\documentclass[12pt]{article}

\newsavebox{\ns}
\newsavebox{\dbrane}
\newsavebox{\dbshort}

\usepackage{epsfig}
\usepackage{amssymb}

\def\appendix{{\newpage\section*{Appendix}}\let\appendix\section%
        {\setcounter{section}{0}
        \gdef\thesection{\Alph{section}}}\section}

\def\be{\begin{eqnarray}}
\def\ee{\end{eqnarray}}

\newcommand{\nn}{\nonumber}
\newcommand\para{\paragraph{}}
\newcommand{\ft}[2]{{\textstyle\frac{#1}{#2}}}
\newcommand{\eqn}[1]{(\ref{#1})}

\newcommand\bomega{\mbox{\boldmath $\omega$}}

\newcommand\bphi{\mbox{\boldmath $\phi$}}

\def\Dslash{\,\,{\raise.15ex\hbox{/}\mkern-12mu D}}
\def\Dbarslash{\,\,{\raise.15ex\hbox{/}\mkern-12mu {\bar D}}}
\def\delslash{\,\,{\raise.15ex\hbox{/}\mkern-9mu \partial}}
\def\delbarslash{\,\,{\raise.15ex\hbox{/}\mkern-9mu {\bar\partial}}}
\def\pslash{\,\,{\raise.15ex\hbox{/}\mkern-9mu p}}
\def\calDslash{\,\,{\raise.15ex\hbox{/}\mkern-12mu {\cal D}}}

\begin{document}
\pagestyle{plain}
\setcounter{page}{1}
\newcounter{bean}
\baselineskip16pt

\begin{titlepage}

\begin{center}
\today
\hfill hep-th/0303151\\
\hfill MIT-CTP-3348 \\

\vskip 1.5 cm
{\Large \bf Mirror Mirror On The  Wall} 
\vskip 0.4cm
{\large (On Two-Dimensional Black Holes and Liouville Theory)}
\vskip 1 cm 
{David Tong}\\
\vskip 1cm
{\sl Center for Theoretical Physics, 
Massachusetts Institute of Technology, \\ Cambridge, MA 02139, U.S.A.\\}

\end{center}

\vskip 0.5 cm
\begin{abstract}
We present a novel derivation of the duality between the two-dimensional 
Euclidean black hole and supersymmetric Liouville theory. We 
realise these $(1+1)$-dimensional conformal field 
theories on the worldvolume of domain walls in a $(2+1)$-dimensional 
gauge theory. We show that there exist two complementary descriptions 
of the domain wall dynamics, resulting in the two mirror conformal field theories. 
In particular, 
effects which are usually attributed to worldsheet instantons are captured by the 
classical scattering of domain walls. 

\end{abstract}

\end{titlepage}

\section{Introduction}

The two-dimensional black hole is a much studied object. It 
was originally introduced in the early '90s as a classical solution of 
two-dimensional string theory \cite{2dbh}. In its Lorentzian form,   
the metric has the structure of the Schwarzchild solution and displays an 
event horizon. In the dark days before D-branes, this background 
was studied as a toy model to explore stringy properties of black 
holes. In these enlightened modern times, attention has focused on the Euclidean 
incarnation  of 
the two-dimensional black hole which has the geometry of a semi-infinite cigar and 
is shown in Figure 1. Of 
particular interest are the various cameo roles that this background plays in 
ten-dimensional string theory:\ it appears as a Calabi-Yau manifold develops 
an isolated singularity \cite{cy};\  in the near-horizon limit of non-extremal 
NS5-branes \cite{ns5};\ and in the double-scaling limit of little string theory 
\cite{dsns5}. At the end of this paper we will extend this list by 
demonstrating how the black hole sigma model arises in the D2-D6 system of IIA 
string theory.
\para
\begin{figure}[htb]
\begin{center}
\epsfxsize=3in\leavevmode\epsfbox{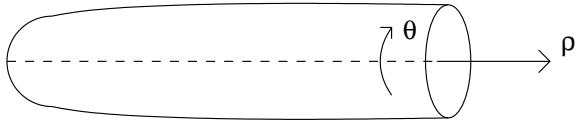}
\end{center}
\caption{The geometry of the two-dimensional Euclidean black hole}
\label{figure}
\end{figure}

The black hole may be described as a 
$SL(2,{\bf R})/U(1)$ coset model at level $k$. The semi-infinite cigar of Figure 1 has a 
metric and dilaton given by
\be
ds_{BH}^2&=&k[d\rho^2+\tanh^2\rho\, d\theta^2] \label{2dbh} \\
\Phi&=&\Phi_0-2\log\cosh\rho
\nn\ee
The geometry is non-singular at the tip $\rho=0$ if the coordinate $\theta$ is 
taken to have periodicity $2\pi$. The non-vanishing 
curvature close to the tip is compensated by the dilaton profile $\Phi$ to 
ensure one-loop conformal invariance. In the asymptotic regime  
$\rho\rightarrow\infty$ the string becomes weakly coupled and propagates  
in a cylindrical geometry of radius $\sqrt{k}$. 
\para
For large $k$ the radius of the asymptotic circle is large and one 
may employ semi-classical techniques to study the $(1+1)$-dimensional 
conformal field theory. For small $k$ one should attempt to find a T-dual description.
It was conjectured by Fateev, Zamolodchikov 
and Zamolodchikov that this T-dual description is a Landau-Ginzburg model with a 
particular potential, known as the sine-Liouville theory \cite{fzz}. 
This duality was used in \cite{kkk} as the starting point in the construction 
of a matrix model for the black hole. 
\para
While much evidence for the duality exists \cite{fzz,kkk}, a proof 
is currently lacking. Progress can be made through the introduction of 
supersymmetry. A  Kazama-Suzuki supercoset construction results in a theory 
with the bosonic background described by equation \eqn{2dbh} and 
${\cal N}=(2,2)$ superconformal symmetry. 
The duality conjecture now states that the 
superconformal theory with target space \eqn{2dbh} is dual to 
super-Liouville theory \cite{dsns5} 
(see also \cite{others} for earlier discussions), 
\be
{\cal L}=\int d^4\theta\,\frac{1}{2k}|Y|^2+\frac{\mu}{2}\left(
\int d^2\theta\,e^{-Y}+{\rm h.c.}\right)
\label{lv}\ee
Here $Y$ is a chiral superfield whose imaginary component has period $2\pi$, 
and $\mu$ is a mass scale.  This action is also accompanied by a linear dilaton. 
The asymptotic regime of this theory ${\rm Re}(Y)\rightarrow\infty$ describes a 
cylinder of radius $1/\sqrt{k}$, as befits the T-dual of the black hole. However, 
the opposite regime ${\rm Re}(Y)<0$ is disfavoured in the Liouville theory by the 
exponential rising potential. This is qualitatively different from the 
cigar metric, where the corresponding regime $\rho<0$ is removed by the 
geometry.
\para
A proof of the equivalence between the theories described by \eqn{2dbh} 
and \eqn{lv} was given by Hori and Kapustin\footnote{Their proof follows 
closely the techniques of \cite{hv}  
prompting them to refer to this duality as ``mirror symmetry''. This, in turn, 
made the title of the current paper sadly unavoidable.} \cite{hk}. 
They realise the black hole background \eqn{2dbh} as the infra-red fixed point 
of a gauged-linear sigma model. One advantage of this construction is that 
it introduces instantons into the picture in the guise of Nielsen-Olesen 
vortices. These appear despite the lack of two-cycles in the target space 
(a related discussion of this phenomenon in the presence of NS5-branes was 
given in \cite{tdualns5}). Hori 
and Kapustin show that the effect of these instantons is to generate the 
superpotential \eqn{lv} upon applying T-duality\footnote{A second 
derivation of this duality was given in \cite{ahkt} by compactifying 
Chern-Simons mirror theories \cite{dt} from three dimensions. In this case  
the instanton effects in two-dimensions are related to one-loop effects in 
three-dimensions.}.
\para
In this paper we present another derivation of the duality, which has a 
very different flavour to previous proofs. Our hope is that these techniques 
may prove useful beyond the situation considered here.
\para
The starting point is a $(2+1)$-dimensional abelian-Higgs model, with a mass 
gap and several isolated vacua. Our interest will be focused on the system of 
$(1+1)$-dimensional domain walls which interpolate between these vacua.  
The basic idea is that there are two different descriptions of the dynamics of the  
domain walls. In the first description, one studies domain walls in the 
classical theory and subsequently quantises their zero modes. 
In the second description, one integrates out the heavy modes in three 
dimensions, and then examines the dynamics of classical domain walls in the 
effective theory. We shall see that these two 
descriptions are precisely the two mirror theories \eqn{2dbh} and 
\eqn{lv}. 
\para
The results of this paper are somewhat reminiscent of ${\cal N}=4$ 
super-Yang-Mills theories in $d=(2+1)$ dimensions \cite{3d}. 
Recall that the instanton effects of three-dimensional gauge theories 
(which are monopole configurations) are captured by the classical dynamics of 
monopoles in a different gauge group\footnote{For a quantitative comparison of 
instanton computations vs. monopole scattering, see \cite{3dinst}}. 
In the present setting, 
we have instanton effects of a two-dimensional gauge theory (which are 
Nielsen-Olesen vortices) once again captured by the classical 
dynamics of solitons: in this case domain walls.
\para
The paper is organised as follows. In the following section we present a 
pair of domain walls which feel no static force. We show that the velocity 
dependent interactions between the walls are described by a non-linear 
sigma-model with target space \eqn{2dbh}. In Section 3 we study the same 
domain walls in an effective, low-energy three-dimensional theory. 
We find that the quantum effects induce a static force between the 
two domain walls which is described by 
the Liouville theory \eqn{lv}. In Section 4, we discuss several further aspects of 
this idea,  including a brane construction in the D2-D6 system of IIA string theory 
and the realisation of other toric sigma 
models --- such as the ${\bf CP}^n$ model --- on the worldvolume of domain walls.  

\section{The Black Hole from Domain Wall Dynamics}

Our starting point is a $d=(2+1)$ dimensional $U(1)$ gauge theory with 
${\cal N}=4$ supersymmetry (8 supercharges). For interesting domain wall 
dynamics, we need three or more isolated vacua, resulting in two or 
more domain walls. We choose the simplest case of three vacua, which 
requires three charged hypermultiplets. The bosonic 
part of the Lagrangian is given by
\be
{\cal L}&=&\frac{1}{4e^2}F_{\mu\nu}F^{\mu\nu}+\frac{1}{2e^2}|\partial{\bphi}|^2
+\sum_{i=1}^3(|{\cal D}q_i|^2+|{\cal D}\tilde{q}_i|^2)
-\sum_{i=1}^3|{\bphi}-{\bf m}_i|^2(|q_i|^2+|\tilde{q}_i|^2)\nn\\
&& -\frac{e^2}{2}
|\sum_{i=1}^3\tilde{q}_iq_i|^2
-\frac{e^2}{2}(\sum_{i=1}^3|q_i|^2-|\tilde{q}_i|^2-\zeta)^2 
\label{lag}\ee
Here $\bphi$ is a triplet of neutral scalar fields which live in the vector 
multiplet. Each hypermultiplet contains two complex scalar fields, $q_i$ and 
$\tilde{q}^\dagger_i$, with charge $+1$ under the $U(1)$ gauge group. In 
three dimensions, each hypermultiplet is assigned a triplet of masses 
${\bf m}_i$.
\para
If the masses ${\bf m}_i$ are distinct, and the FI parameter is strictly
 positive  $\zeta>0$,  then the theory has three isolated, massive vacua, given by
\be
{\rm Vacuum\ }i:\ \ \ \bphi={\bf m}_i\ \ ,\ \ 
|q_j|^2=\zeta\,\delta_{ij}\ \ ,\ \ 
|\tilde{q}_j|^2=0\ \ \ \ \ \ \ \ i=1,2,3
\label{vac}\ee
Let us examine the vacuum physics. In each ground state, supersymmetry is unbroken and 
the theory is in a Higgs phase. Together with the broken $U(1)$ gauge symmetry, the 
theory also exhibits an unbroken $U(1)_F^2$ flavour symmetry, acting on the $q_i$ and 
$\tilde{q}_i^\dagger$. All three vacua have a mass gap, with the 
lightest excitations depending on the ratio 
$k\sim \zeta/\Delta{\bf m}_i$ where $\Delta {\bf m}_i$ are the mass splittings. 
In the limit $k\rightarrow \infty$, hypermultiplet modes become massless, reflected 
by the appearance of a Higgs branch. In contrast, as $k\rightarrow 0$, the 
vector multiplet becomes light and a Coulomb branch of vacua appears. 
The parameter $k$ will be defined more precisely below, and plays an important role in 
our story.

\subsubsection*{\it Classical Domain Wall Dynamics}

The system of $\ft12$-BPS domain walls that interpolate between these 
isolated vacua has been studied in \cite{kinky,dw1,dw,keith}. 
The tension of a domain wall interpolating from  
the $i^{\rm th}$ to the $j^{\rm th}$ vacuum is given by 
$T_{ij}=\zeta|{\bf m}_j-{\bf m}_i|$. This suggests that something 
special may happen when we choose the mass triplets to be co-linear  
${\bf m}_i=(m_i,0,0)$ with, say, $m_{i+1}>m_i$. For in this case, 
we have
\be
T_{13}=T_{12}+T_{23}
\nn\ee
which is a necessary, although not sufficient, condition for there to be 
no classical force between a domain wall 
interpolating from the $1^{\rm st}$ to the $2^{\rm nd}$ vacuum, and a 
domain wall interpolating from the $2^{\rm nd}$ to the $3^{\rm rd}$ vacuum. 
In fact this energetic reasoning is correct and there do exist static 
solutions corresponding to domain walls with arbitrary separation $R$ as 
shown in Figure 2. 
\begin{figure}[htb]
\begin{center}
\epsfxsize=4in\leavevmode\epsfbox{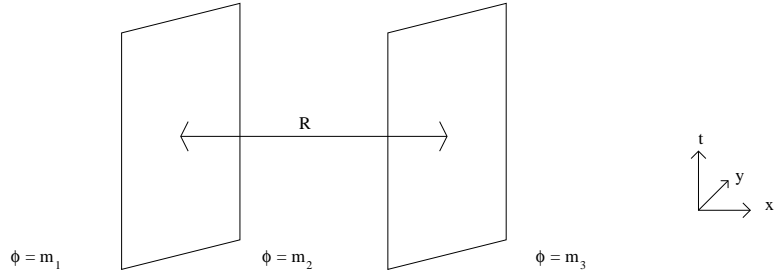}
\end{center}
\caption{Two domain walls. (No, really). The separation $R$ is a modulus.}
\label{figure2}
\end{figure}
\para
To see that this is the case, we examine the zero modes structure 
of solutions to the Bogomoln'yi equations describing these domain walls. 
These equations, first derived in \cite{kinky}, involve only $\bphi=(\phi,0,0)$,  
\be
\partial\phi&=&e^2(\sum_{i=1}^3|q_i|^2-\zeta) \nn\\
{\cal D}q_i&=&(\phi-m_i)q_i
\label{bog}\ee
where $\partial\phi$ denotes the derivative of $\phi$ with respect to the 
spatial coordinate $x$ transverse to the domain wall, and 
${\cal D}q=\partial q-iA_xq$ is the covariant derivative.
In the strong coupling limit $e^2\rightarrow\infty$, it was shown in 
\cite{dw1,dw} that these equations do admit a moduli space of 
solutions, with one of the collective coordinates having the 
interpretation of the separation between the walls. At finite 
$e^2$, an index theory computation was performed by Lee \cite{keith}, 
revealing that solutions to \eqn{bog} 
indeed have the relevant zero modes. 
\para
Let us describe these zero modes in 
more detail. Each of the walls has two collective coordinates. 
One of these is simply the position of the domain wall, while the second 
is an internal, periodic degree of freedom which comes from acting 
on the domain wall solution with one of the $U(1)_F$ flavor symmetries 
\cite{abtown}. These are analogous to the collective coordinates 
of monopoles that arises from large gauge transformations. Our system 
of two domain walls therefore has $4$ collective coordinates, corresponding 
to their center of mass, their separation, and $2$ internal phases. 
Furthermore, there are $8$ fermionic zero modes, $4$ of which arise 
from broken supersymmetry while the remaining ones are guaranteed by the 
unbroken ${\cal N}=(2,2)$ supersymmetry preserved on the worldvolume 
of the domain walls.
\para
The low-energy dynamics of the domain walls is thus described 
as a $d=(1+1)$-dimensional sigma model with 
${\cal N}=(2,2)$ supersymmetry, and with target space given by the 
domain wall moduli space ${\cal M}$. The simplest case to consider is when 
the two domain walls have equal tension. This occurs when the mass parameters 
are given by $m_i=(-M/2,0,M/2)$ which means that $T_{12}=T_{23}=\zeta M/2$. 
In this case, the moduli space of two domain walls has the structure \cite{dw1},
\be
{\cal M}={\bf R}\times\frac{{\bf S}_1\times\tilde{\cal M}}{{\bf Z}_2}
\label{moduli}\ee
where the ${\bf R}$ factor is 
parameterised by $r$, the center of mass of the domain 
walls, while the ${\bf S}^1$ factor is parameterised by $\tau\in[0,2\pi)$, 
the overall phase of the domain walls. All interesting information 
about the dynamics of the domain walls is contained in the 
relative moduli space $\tilde{\cal M}$. 
This two-dimensional manifold is parameterised by a variable 
$R\in(-\infty,\infty)$ and the relative phase $\theta\in[0,2\pi)$.
For large values, $R$ is equal to the separation between the domain walls. 
However at distances less than $(\zeta M)^{-1/2}$, when the domain wall cores overlap, 
this interpretation breaks down as is obvious from the fact that $R$ 
takes negative values. The ${\bf Z}_2$ action acts as 
$\chi\rightarrow\chi+\pi$ and $\theta\rightarrow\theta +\pi$. 
The metric on ${\cal M}$ was calculated in \cite{dw} and is given by,
\be
ds_{DW}^2=k\left[dr^2+d\tau^2+e^{\psi_{DW}(R)}\left(\frac{M^2}{16}dR^2+
d\theta^2\right)\right]\label{metric}\ee
The overall scale of the metric is
\be
k=\frac{2\zeta}{M}
\label{k}\ee
and will soon be identified with the level $k$ of the coset space construction 
of the 2D black hole \eqn{2dbh}. To compare to the black hole metric, it will 
prove convenient to measure the separation between domain walls in terms of 
the dimensionless quantity,
\be
u=\frac{RM}{4}
\nn\ee
All information about the classical scattering of domain walls is encoded 
within the smooth, real, and somewhat ugly function  $\psi_{DW}(u)$, given 
by \cite{dw}
\be
e^{\psi_{DW}(u)}=\frac{e^{2u}}{e^{4u}-4}\left(e^{2u}-\frac{4}{\sqrt{4-e^{4u}}}
\cos^{-1}(e^{2u}/2)\right)
\nn\ee
Although somewhat hidden in these coordinates, the relative moduli space 
of the domain walls does look like the  semi-infinite cigar of Figure 1. 
This is obvious in the limit $u\rightarrow\infty$, where 
$e^{\psi(u)}\rightarrow 1$, so that the moduli space becomes cylindrical. 
It is less obvious that the moduli space is cigar-like at 
$u\rightarrow -\infty$, but this is made more plausible by noting that 
the point at $-\infty$ is at finite affine distance 
in the metric \eqn{metric}. A suitable coordinate transformation brings 
this point to finite parametric distance, where it can be checked that the 
tip of the cigar is smooth \cite{dw}. Rather than take this route, we will 
instead bring the black hole metric into the form of \eqn{metric}. 
Define $\sinh\rho=e^u$, so that the metric \eqn{2dbh} becomes,
\be
ds^2_{BH}=ke^{{\psi}_{BH}(u)}\left(du^2+d\theta^2\right)
\label{bh}\ee
where 
\be
e^{\psi_{BH}(u)}=\frac{e^{2u}}{1+e^{2u}}
\nn\ee
Note that both $\psi_{DW}(u)$ and $\psi_{BH}(u)$ have the same 
asymptotic behaviour,
\be
\psi(u)\rightarrow\left\{\begin{array}{ll} 0 & u\rightarrow +\infty \\
2u & u\rightarrow -\infty\end{array}\right.
\label{bc}\ee
Away from this asymptotic regime, the metrics differ. 

\subsubsection*{\it Quantum Domain Wall Dynamics}

Neither the domain wall metric \eqn{metric}, nor the black hole metric 
\eqn{bh}, are Ricci flat. This non-zero curvature gives a one-loop contribution 
to the beta-function of the form $\beta_{ij}=-R_{ij}/2\pi$. In the absence of 
any other sources, the metric therefore changes with 
scale $t$. For target spaces of the cigar form the RG equations read,
\be
ke^\psi\frac{\partial\psi(u,t)}{\partial t}=\frac{1}{4\pi}
\frac{\partial^2\psi(u,t)}{\partial u^2}
\label{rgflow}\ee
In the two-dimensional black hole solution \eqn{2dbh}, the term on the 
right-hand side is canceled by the contribution from the dilaton. 
As shown by Hori and Kapustin \cite{hk}, one can mimic the 
dilaton contribution by assigning an anomalous transformation to the 
target space coordinates under Weyl rescaling. The basic idea is that 
the curvature at the tip of the cigar causes the cigar to shrink. To keep 
things looking conformal, we should run along with this shrinking tip. 
Under the condition that the function $\psi(u)$ obeys the boundary conditions 
\eqn{bc}, then it can be shown \cite{hk} that the existence a conformal fixed point 
requires the anomalous Weyl transformation, 
\be
u\rightarrow u + \frac{t}{2\pi k}
\nn\ee
From the perspective of the domain walls, as an observer examines the physics 
on larger and larger distance scales ($t\rightarrow\infty$) she must increase 
the distance between the domain walls in an attempt to keep things looking 
the same. 
Writing $\psi$ as a function of the scale invariant quantity $v=u-t/2\pi$, 
the RG equations become
\be
ke^\psi\frac{\partial\psi(v,t)}{\partial t}=\frac{1}{4\pi}
\frac{\partial^2\psi(v,t)}{\partial v^2}+\frac{e^\psi}{2\pi}
\frac{\partial\psi (v,t)}{\partial v}
\nn\ee
It is simple to check that the unique conformal solution ($\dot{\psi}=0$) to 
this equation is the black hole metric \eqn{bh}, where second term on the 
right-hand-side has played the role of the dilaton. We are now in a position to see 
how the domain wall metric \eqn{metric} 
evolves under the RG group. It is a simple matter to check numerically that 
the domain wall metric does indeed flow towards the conformal black 
hole metric \eqn{bh} in the infra-red. 
\para
The one-loop approximation to the beta-function requires large $k$, 
and our derivation of the black hole conformal field theory is therefore 
valid only in this limit. It is possible to go beyond this approximation using a 
further result of Hori and Kapustin \cite{hk}: there is a unique SCFT with the asymptotic 
behaviour and symmetries of the black hole and central charge,
\be
c=3\left(1+\frac{2}{k}\right)
\nn\ee
which is indeed the central charge of the Kazama-Suzuki $SL(2,{\bf R})/U(1)$ coset 
construction at level $k$.
Thus to establish the connection to the black hole SCFT for all values of $k$, we 
must determine the infra-red central charge of the domain wall dynamics. This remains 
an open problem. 
 
\subsubsection*{\it Losing Supersymmetry}

The above discussion has, for the most part, not relied on the supersymmetry of 
the three-dimensional field theory. Of course, the three dimensional fermions 
supply the zero modes which ensure that we end up with  
${\cal N}=(2,2)$ supersymmetry on the domain wall worldvolume, but these fermions affect 
neither the classical metric \eqn{metric} nor the RG flow equations \eqn{rgflow}. 
We could happily start from the three-dimensional bosonic Lagrangian 
\eqn{lag}, omitting the fermions, and arrive at the same low-energy description of the 
domain wall
dynamics in terms of the bosonic CFT on the black hole background \eqn{2dbh}. The 
only deviation from the formulas above is that the central charge is now expected to 
be \cite{2dbh}
\be
c_{bosonic}=\frac{3k}{k-2}-1
\nn\ee
Curiously, for the purely bosonic black hole, the coset construction only makes 
sense for $k>2$. From our perspective, this requires that classically $\zeta>M$. It would 
be interesting to understand this restriction in terms of phase transitions 
of the three-dimensional field theory. When quantum effects are taken into account, 
it seems plausible that the bosonic 
theory exits the gapped Higgs phase and enters a massless phase around 
$\zeta\sim M$.

\section{Liouville Theory from Domain Wall Dynamics} 

In this section we present a different description of the domain wall dynamics, 
valid when $k=2\zeta/M\ll 1$. To see what this new description may involve, 
note that in the limit $k\rightarrow 0$ the vector multiplet fields become light. 
This suggests that we should integrate out the hypermultiplets, leaving behind 
an effective 
Lagrangian for the photon, the triplet of neutral scalar fields $\bphi$, and their 
fermionic partners. This technique of examining theories for different values 
of $k$ was employed in the context of supersymmetric quantum mechanics in \cite{denef}. 
\para
One important point is that in three 
dimensions we can dualise the photon in favour of 
a periodic scalar field $\sigma$ defined as $F={}^\star d\sigma$. 
The low-energy effective action is given by a massive hyperk\"ahler sigma-model 
preserving ${\cal N}=4$ supersymmetry, with the bosonic interactions described by
\cite{is}\be
{\cal L}=\ft12 H(\bphi)\,|\partial\bphi|^2 + 2 H(\bphi)^{-1}
(\partial\sigma+\bomega\cdot\partial\bphi)^2-\ft12\zeta^2H(\bphi)^{-1}
\label{intout}\ee
where $\bomega$ is defined as $\nabla\times\bomega =\nabla H$, and 
\be
H(\bphi)=\frac{1}{e^2}+\sum_{i=1}^3\frac{1}{|\bphi-{\bf m}_i|}
\label{harm}\ee
The first two terms in \eqn{intout} describe a sigma-model with the 
target space given by the Gibbons-Hawking metric with three centers. 
Notice in particular the interaction term between the dual photon $\sigma$ 
and the scalars $\bphi$ which arises from integrating out fermions in the 
hypermultiplet: we shall return to the importance of this coupling shortly. 
The final term in \eqn{intout} is a potential, lifting the Coulomb branch. It 
is  the unique potential allowed by ${\cal N}=4$ supersymmetry \cite{agf}.
The Lagrangian \eqn{intout} preserves the vacuum structure of the classical 
theory \eqn{vac}, with three supersymmetric ground states given by
\be
{\rm Vacuum\ }i:\ \ \ \bphi={\bf m}_i\ \ \ \ \ \ \ \ \ \ i=1,2,3
\nn\ee
In each of these vacua, the kinetic terms in \eqn{intout} are singular, but 
this is simply a coordinate singularity and may be eliminated by a field 
redefinition in which $\sqrt{\phi}$ is treated as the canonically normalised field.

\subsubsection*{\it Classical Domain Wall Dynamics}

Solitons in massive sigma-models of this type have been much studied 
in the literature \cite{abtown,dw1,dw,dw2,dw3,dw4,dw5}. As for the classical theory of the 
previous section, the BPS domain walls have tension 
$T_{ij}=\zeta|{\bf m}_i-{\bf m}_j|$. Restricting 
once again to co-linear masses, ${\bf m}_i=(m_i,0,0)$, the 
BPS equations describing domain walls are given by
\be
\partial\phi -\zeta\, H(\phi)^{-1}&=&0\label{bog1}\\
\partial\sigma+\bomega\cdot\partial\bphi &=&0 
\label{bog2}\ee
Let us start by examining the first of these equations. We are 
interested in solutions which interpolate between the first vacuum at 
$\bphi={\bf m}_1$ and the third vacuum at $\bphi={\bf m}_3$. 
However, equation \eqn{bog1} involves only a single scalar field 
$\bphi=(\phi,0,0)$ and it is well known that such systems admit solutions 
only for domain walls interpolating between neighbouring vacua. There 
is no solution interpolating between the first and third vacua. One could 
of course attempt to construct an approximate solution by considering a 
configuration of two, well-separated domain walls. But, as we shall 
review in detail below, such a configuration results in a repulsive 
force between 
the two domain walls. This force occurs despite the fact that each 
wall preserves (the same) half of the eight supersymmetries in three 
dimensions and therefore provides a counterexample to the commonly 
stated maxim that BPS necessarily implies no force (for a 
discussion of this system, see \cite{dw1}).
\para
The above results are in stark contrast to the previous section where we 
saw that there exist solutions in the classical theory corresponding to 
domain walls with arbitrary separation. In that case, the existence of these 
solutions resulted in a description of the quantum dynamics of the domain walls 
in terms of a sigma-model. In the present case, the existence of a potential 
means that the description of the domain wall dynamics will be in the form of a 
Landau-Ginzburg theory. In the remainder of this 
section, we shall calculate the force experienced by two well-separated domain 
walls, and show that their dynamics is governed by the Liouville theory 
\eqn{lv}.

\subsubsection*{\it The Force Between Domain Walls}

A technique for computing the low-energy dynamics describing the repulsive force 
between domain walls was presented long ago by Manton \cite{manton}, and discussed 
more recently in \cite{portown}. Manton's method is a beautifully direct and simple 
approach to the problem, and proceeds as follows: One firstly constructs an approximate 
solution to the equations of motion consisting of a superposition of two well-separated 
domain walls, suitably patched together in the middle. One then examines how this 
configuration evolves in time under the full 
second order equations of motion and calculates the acceleration of the part of the 
field configuration describing just one of two kinks. The magic of the approach lies in the 
fact that this acceleration is independent of exactly where one chooses to do the patching. 
Let us now see in detail how to apply Manton's 
procedure to our situation.
\para
To start, we wish to construct an approximate solution consisting of two 
well-separated domain walls. We will be aided in this task by knowing the 
solution for a single domain wall so let us begin with this. 
For clarity, we will keep the masses $m_i$ arbitrary, subjected only 
to $\sum_im_i=0$. Only later will we restrict to domain walls of equal tension. 
Consider first the kink which interpolates between the first and second 
vacua, with tension $T_{12}=\zeta(m_2-m_1)$.  Throughout space the field $\phi$ lies within 
the range $m_1\leq\phi\leq m_2<m_3$, and the domain wall equations therefore read,
\be
\begin{array}{lll}
\hspace{.1in}
{\rm kink\ }1\rightarrow 2:\hspace{1in}&
\lefteqn{\partial\phi=\zeta\left(\frac{1}{e^2}+\frac{1}{\phi-m_1}+\frac{1}{m_2-\phi}
+\frac{1}{m_3-\phi}\right)^{-1}}&\hspace{3in}\ 
\end{array}
\nn\ee
For finite $e^2$, one can reduce this to an algebraic equation for $\phi$. 
However, in the limit $e^2\rightarrow\infty$, an explicit solution is 
known \cite{abtown}. If we take the domain wall to be centered at $x=x_1$, then the kink 
profile $\phi_{12}(x-x_1)$ is given by the solution to the quadratic equation,
\be
\begin{array}{lll}
\hspace{.1in}
{\rm kink\ }1\rightarrow 2:\hspace{1in}&
\lefteqn{e^{\zeta(x-x_1)}(m_2-\phi_{12})(m_3-\phi_{12})-m_3(\phi_{12}-m_1)=0}&
\hspace{3in}\ 
\end{array}
\nn\ee
where, in order to get the correct boundary conditions, we must take the 
negative square root. It will prove useful to have the asymptotic form 
of the profile to the far right of the kink. As $(x-x_1)\rightarrow \infty$, we have
\be
{\rm kink\ }1\rightarrow 2:\hspace{.75in}
\phi_{12}\ \rightarrow\ m_2-m_2\frac{m_2-m_1}{m_3-m_2}e^{-\zeta(x-x_1)}+{\cal O}
(e^{-2\zeta(x-x_1)})
\label{ex1}\ee
There is a similar story for the kink interpolating between the second and 
third vacua with tension $T_{13}=\zeta(m_3-m_2)$. This time 
$m_1<m_2<\phi<m_3$, so that the the equation of motion reads,
\be
\begin{array}{lll}
{\rm kink\ }2\rightarrow 3:\hspace{.8in}&
\lefteqn{\partial\phi=\zeta\left(\frac{1}{e^2}+\frac{1}{\phi-m_1}+\frac{1}{\phi-m_2}
+\frac{1}{m_3-\phi}\right)^{-1}}&\hspace{3.2in}\ 
\end{array}
\nn\ee
The profile of the kink positioned at $x=x_2$ is given by the solution to 
the quadratic equation,
\be
\begin{array}{lll}
{\rm kink\ }2\rightarrow 3:\hspace{.8in}&
\lefteqn{e^{-\zeta(x-x_2)}(\phi_{23}-m_1)(\phi_{23}-m_2)+m_1(m_3-\phi_{23})=0}
&\hspace{3.2in}\ 
\end{array}
\nn\ee
This time we will be interested in the 
asymptotic regime far to the left of this kink. 
As $(x-x_2)\rightarrow -\infty$, the profile looks like
\be
\begin{array}{lll}
{\rm kink\ }2\rightarrow 3:\hspace{.8in}&
\lefteqn{\phi_{23}\ \rightarrow\ m_2-m_1\frac{m_3-m_2}{m_2-m_1}e^{\zeta(x-x_2)}+{\cal O}
(e^{2\zeta(x-x_2)})}&\hspace{3.2in}\ 
\end{array}
\label{ex2}\ee
We now wish to construct a configuration which looks like two far-separated 
domain walls. An obvious guess is to fix $x_2-x_1$ to be large, and 
write $\phi_{13}(x)=\phi_{12}(x-x_1)+\phi_{23}(x-x_2)-m_2$, which has the 
correct boundary conditions. However, one can check that this configuration 
is actually singular where they join because of the coordinate singularity 
in the target space of \eqn{intout} at the point $\phi=m_2$. The correct 
procedure to patch the two kinks together is to transform to a basis of 
fields which have canonical kinetic term at the point where the two 
kinks join $\phi=m_2$. This is achieved by setting
\be
f(\phi_{13}(x)-m_2)=f(\phi_{12}(x-x_1)-m_2)+f(\phi_{23}(x-x_2)-m_2)
\label{f}\ee
where the function $f(\phi-m_2)$ is defined as
\be
f(\phi-m_2)=\left\{\begin{array}{ll}\sqrt{\phi-m_2} \ \ \ \ \ \ \ \ 
& \phi\geq m_2 \nn\\
-\sqrt{m_2-\phi} & \phi< m_2\end{array}\right.
\nn\ee
The configuration $\phi_{13}(x)$ described by \eqn{f} is our 
initial condition for a profile describing 
two well-separated domain walls. The next step is to understand how 
this configuration evolves under the equations of motion of the system. 
After consistently truncating to the field of interest $\phi$, the 
equations of motion derived from \eqn{intout} read,
\be
\ft12\frac{\partial H}{\partial\phi} (\partial\phi)^2
+H\partial^2\phi-\ft12\zeta^2H^{-2}
\frac{\partial H}{\partial\phi}=0
\label{eom}\ee
We will consider the change in momentum experienced by the first kink. 
The momentum $P$ of the field between the point $x=a$ and the point 
$x=b$ is given by $T_{0x}$ component of the energy-momentum tensor,
\be
P=-\int_{a}^b dx\ H(\phi)\dot{\phi}\phi^\prime
\nn\ee 
from which we can compute rate of change of the momentum, 
\be
\dot{P}&=&-\int_a^bdx\ \frac{\partial H}{\partial \phi}\dot{\phi}^2\phi^\prime
+H\ddot{\phi}\phi^\prime+H\dot{\phi}\dot{\phi}^\prime \nn\\
&=&\left[-\ft12H(\dot\phi^2+\phi^{\prime\, 2})+\ft12\zeta^2H^{-1}\right]^b_a
\label{pdot}\ee
where we have made use of the equation of motion \eqn{eom} in the second 
line to express $\dot{P}$ as a total derivative. We now evaluate 
\eqn{pdot} on our two-kink configuration \eqn{f}, with the initial 
conditions $\dot{\phi}_{13}=0$. Since we wish to compute the acceleration of 
the part of the field corresponding to the first kink, we set 
$a\rightarrow -\infty$ while choosing $b$ to lie somewhere between 
the two domain walls. We require only that both $(b-x_1)$ and $(x_2-b)$ are large. 
We then expand the domain wall configuration \eqn{f} to leading order, 
using the expressions  for the two kink solutions \eqn{ex1} and \eqn{ex2}. 
Upon substituting these expressions into \eqn{pdot}, the calculation 
offers up two minor miracles: the final answer is independent of 
$b$, and independent of the ubiquitous ratio $(m_2-m_1)/(m_3-m_2)$. We find,
\be
\dot{P}=-2\zeta^2\sqrt{-m_1m_3}e^{-\zeta(x_2-x_1)/2}
\label{yes}\ee

\subsubsection*{\it The Low-Energy Effective Lagrangian}

Using Manton's method, we have computed the force experienced by the 
separated domain walls \eqn{yes}. Since each action has an equal and 
opposite reaction [New3], $\dot{P}=T_{12}\ddot{x}_1=-T_{12}\ddot{x}_2$, from which we find the 
rate 
of change of the separation $R$ between the kinks,
\be
\ddot{R}=\ddot{x}_2-\ddot{x}_1=2\zeta\sqrt{-m_1m_3}\frac{(m_3-m_1)}{(m_2-m_1)(m_3-m_2)}
e^{\zeta(x_2-x_1)/2}
\label{eom1}\ee
Newton's first law is also useful, giving us the formula for the center of mass 
acceleration $\ddot{r}=\ft12(\ddot{x}_2+\ddot{x}_1)=0$. 
At this point, we specialise to two domain walls of equal tension by setting the 
bare masses equal to $m_i=(-M/2,0,M/2)$ as in Section 2. 
The equation of motion \eqn{eom1} for the domain wall then follows 
from the $(1+1)$-dimensional effective action
\be
{\cal L}_{\rm position}=k\left[(\partial r)^2+\frac{M^2}{16}(\partial R)^2\right]-2
\zeta Me^{-\zeta R/2}
\nn\ee
where we have covariantised the kinetic term 
to ensure the Lagrangian is Lorentz invariant along the domain wall. The 
overall normalisation of this Lagrangian is in agreement with \eqn{metric} 
and is correct for the motion of a two domain walls, each of tension 
$T=M\zeta/2$. 
\para
So far we have considered collective coordinates associated to the position 
of the domain wall. As in the previous section, there exist further 
periodic, internal degrees of freedom for each wall. 
These arise from the second Bogomoln'yi equation \eqn{bog2}. 
To account for these, we must firstly rewrite the triplet of scalars $\bphi$ in 
polar coordinates,
\be
\bphi=(\phi,\xi\cos\chi,\xi\sin\chi)
\nn\ee
in terms of which the second Bogomoln'yi equation \eqn{bog2} becomes
\be
\sigma^\prime+\sum_{i=1}^3\frac{\phi-m_i}{\sqrt{\xi^2+(\phi-m_i)^2}}\chi^\prime
=0
\label{second}\ee
where both $\sigma$ and $\chi$ have period $2\pi$. 
Thus, in the background of the first kink which interpolates from the first to 
second vacua, we have 
\be
\begin{array}{ll}
\hspace{.1in}{\rm kink\ }1\rightarrow 2:\hspace{.8in}&\sigma^\prime-\chi^\prime=0
\end{array}
\nn\ee
while, in the background of the second kink, interpolating from the 
second to the third vacua, 
\be
\begin{array}{ll}
\hspace{.1in}
{\rm kink\ }2\rightarrow 3:\hspace{.8in}&\sigma^\prime+\chi^\prime=0
\end{array}
\nn\ee
The relative minus signs between these two equations ensures the 
following, crucial fact: in the 
background of the two kink configuration \eqn{f} both the value of 
$\sigma$ and the value of $\chi$ remain good collective coordinates, and 
neither field appears in the potential. Their 
kinetic terms in the background \eqn{f} may be determined from the 
Lagrangian \eqn{intout} and are given by,
\be
{\cal L}_{\rm phase}=\frac{4}{k}\left((\partial \sigma)^2+(\partial \chi)^2\right)
\nn\ee
Comparing to the effective domain wall action that we found in Section 2  
\eqn{metric}, we see that 
$\sigma$ and $\chi$ are related to $\tau$ and $\theta$ respectively by T-duality. 
The final bosonic Lagrangian is given by ${\cal L}_{\rm position}+{\cal L}_{\rm phase}$, 
from which we extract the part concerned with the relative motion of the 
domain walls involving the fields $R$ and $\chi$. An important observation is 
that this low-energy theory has a natural complex structure,
\be
Y=\zeta R/4 +i\chi
\nn\ee
in terms of which the relative domain wall dynamics are described by,
\be
{\cal L}_{\rm relative}=8\left(\frac{1}{2k}|\partial Y|^2-\frac{\zeta M}{4}
|e^{-Y}|^2\right)
\label{answer}\ee
Here the overall coefficient $8$ has been factored out to ensure that the 
normalization of the periodic coordinate agrees with the bosonic terms from 
the Liouville theory \eqn{lv}. Having matched these kinetic terms, any non-trivial 
comparison with the Liouville model lies 
with the potential. The overall scale of the potential simply determines the 
subtraction point $\mu$ of the two-dimensional theory in terms of three 
dimensional parameters: $\mu=\zeta M/4$. This leaves us with only the coefficient 
of the exponent to check. Comparing the potential of \eqn{answer} with the 
superpotential of \eqn{lv}, we see that the repulsive force between the two 
domain walls is indeed described by the Liouville theory \eqn{lv}.

\subsubsection*{\it Losing Supersymmetry}

Once again, we could examine how the above argument fares in the 
absence of supersymmetry. The conjecture is that if we start from 
the three-dimensional purely bosonic Lagrangian \eqn{lag}, then 
the relative dynamics of 
domain walls are described by the sine-Liouville theory with 
Lagrangian \cite{fzz,kkk}
\be
\tilde{{\cal L}}=\frac{1}{k-2}(\partial y)^2+\frac{1}{k}(\partial\chi)^2
-\mu^2e^{-y}\cos\chi
\label{sinelv}\ee
where $y$ is related to the separation of the walls. There are two key differences 
between this Lagrangian and the super-Liouville theory. The first is that 
shifts of $\chi$ are no longer a symmetry in the bosonic case. The second is
that  the model is unstable for $k<2$, mimicking the fact that the coset 
construction of the black hole only makes sense for $k>2$.
\para 
Can we reproduce \eqn{sinelv} from domain wall dynamics? Unfortunately, 
and in contrast to Section 2, supersymmetry played a vital role in 
determining the domain wall dynamics in this Section. 
Without the strong holomorphic (in fact, hyperK\"ahler) restrictions 
imposed by supersymmetry on the low-energy three-dimensional effective 
action \eqn{intout}, we have 
no control over the physics near 
the vacua $\phi=m_i$. Nonetheless, it is interesting to note that at least the 
symmetries coincide with the sine-Liouville potential. To see this, recall 
that the existence of two periodic collective coordinates $\sigma$ and 
$\chi$ in the supersymmetric case 
can be traced to the second term in \eqn{second} which  
first appeared as a $(\partial\sigma+\bomega\cdot\partial\bphi)^2$ coupling in 
\eqn{intout}. In the original variables of the $U(1)$ gauge field, this term 
takes the form 
\be
\epsilon_{\mu\nu\rho}F_{\mu\nu}\,\bomega\cdot\partial_\rho\bphi
\nn\ee
which arises at one-loop from a triangle graph with fermions running in 
the loop (as is clear from the tell-tale presence of the $\epsilon$-symbol). 
We see therefore that the three-dimensional fermions provide the delicate 
mechanism by which the relative internal degree of freedom $\chi$ preserves its 
shift symmetry. If we remove the fermions in three dimensions, there is 
nothing to prevent the appearance of $\chi$ in the bosonic potential.

\section{Discussion}

\begin{figure}[htb]
\begin{center}
\epsfxsize=5in\leavevmode\epsfbox{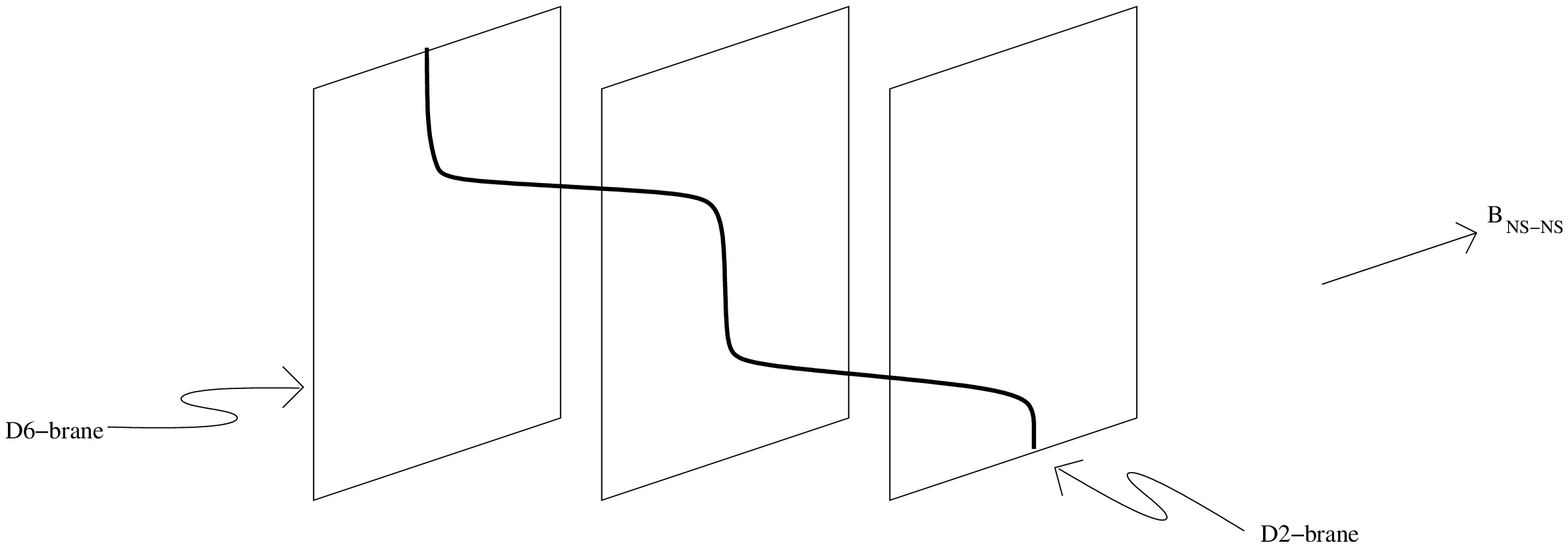}
\end{center}
\caption{Two domain walls in the D2-D6 system. In contrast to other 
pictures, the sheets represent the D6-branes. The vacua of the field 
theory occur when the D2-brane lies vertically within a D6-brane. The 
domain walls correspond to places where 
the D2-brane interpolates horizontally between 
the D6-branes.}
\label{figure4}
\end{figure}

In this, final section, I would like to discuss a few variations on the 
theme. We will start with a description of how these domain walls can appear 
in brane dynamics. This adds one further entry to the list of occurrences of 
the two-dimensional Euclidean black hole in critical superstring theory. We 
will then discuss further generalisations of the idea within field theory and 
add some speculations.

\subsubsection*{\it A Brane Construction}

The brane construction for domain walls has been discussed several times in 
the literature \cite{paul,kinky,dw1} and the only new point here is the 
relationship to the two-dimensional black hole described in Section 2. 
The set-up starts with the familiar D2-D6 system in IIA string theory 
(the M-theory lift was described in \cite{paul}). 
The low-energy interacting modes on a single D2-brane 
in the background of 3 parallel, separated D6-branes are described by the 
Lagrangian \eqn{lag}, where 
the FI parameter $\zeta$ is induced by a background NS-NS B-field. This 
B-field induces an attractive force between the D2-brane and D6-branes, 
resulting in three supersymmetric vacuum states in which the D2-branes lies 
within one of the D6-branes. The supersymmetric domain wall solutions of the 
field theory appear when the D2-brane interpolates from one D6-brane to another 
as shown in Figure 3. When the separation between the domain walls is 
small compared to the B-field then $k\gg 1$ and the dynamics of the 
domain walls is naturally given in terms of the weakly coupled black hole 
sigma model \eqn{2dbh}. When the separation grows, the weakly coupled 
description is in terms of the supersymmetric Liouville theory \eqn{lv}.

\subsubsection*{\it Generalisations}

There is a natural generalisation of the discussion in this paper to 
three dimensional theories with $(N+1)$ hypermultiplets. If each 
of these has a distinct mass ${\bf m}_i$, then there are $(N+1)$ vacua 
and one may consider the dynamics of domain walls interpolating from the 
first to the last. This system was studied in detail in \cite{dw,keith}. 
For co-linear masses ${\bf m}_i=(m_i,0,0)$, it is known that the most 
general solution has $2N$ collective coordinates which have 
the interpretation of the position and internal phase of $N$ 
component domain walls. The moduli space has the structure,
\be
{\cal M}_N={\bf R}\times\frac{{\bf R}\times\tilde{M}_{N-1}}{\cal G}
\nn\ee
where the first two ${\bf R}$ factors parameterise the center-of-mass and 
overall phase of the soliton respectively. The relative kink moduli space 
$\tilde{\cal M}_{N-1}$ has complex dimension $(N-1)$ and is endowed with a 
K\"ahler metric. The precise form of this metric is unknown for $N>2$, 
although it can be shown to have $(N-1)$ holomorphic isometries which are 
inherited from the $U(1)^N$ flavour symmetry of the three-dimensional 
gauge theory.  The quotient by the discrete group ${\cal G}$ acts only on the 
toric fibers and has no fixed points. Thus, $\tilde{\cal M}_{N-1}$ is a 
smooth, non-compact, 
toric K\"ahler 
manifold, and the dynamics of the classical domain walls is described 
by the ${\cal N}=(2,2)$, $d=1+1$ non-linear sigma-model on $\tilde{\cal M}_{N-1}$. 
\para
In the other description of domain walls, one integrates out the $(N+1)$ 
hypermultiplets, and restricts to the Coulomb branch where the low-energy 
dynamics is given by \eqn{intout} with the harmonic function $H(\phi)$ 
given by the natural generalisation of \eqn{harm} to include $(N+1)$ terms. 
In this case, there is again a force between domain walls, resulting in a 
Landau-Ginzburg description of the dynamics. We leave the exact computation 
of this potential for future work, but the resulting theory is expected to be 
the Landau-Ginzburg mirror of the toric-sigma model described in \cite{hv}. 
\para
Let us mention a limit in which other, more familiar, sigma models 
can be realised on the domain wall worldvolume. Consider, for example, the 
situation with $4$ hypermultiplets so that we are studying a system of 
$3$ domain walls as shown in Figure 4. The tension of the $i^{\rm th}$ 
domain wall is given by $T_i=\zeta(m_i-m_{i+1})$. We can consider the 
limit in which $m_1\rightarrow -\infty$ and $m_4\rightarrow +\infty$ 
with $m_2$ and $m_3$ kept finite. This ensures that the first and 
third domain wall become very heavy and hence static, fixed at some 
distance $L$. Meanwhile, the middle domain wall is free to move but 
restricted to lie between the outlying domain walls which now act as bookends. 
The moduli space of this system therefore takes the form of an interval 
parameterised by $R\in[0,L]$. Added to this, we have the internal phase 
$\theta\in {\bf S}^1$ associated with the middle domain wall. Since 
$\tilde{\cal M}_{2}$ is smooth, the submanifold describing the interactions 
of the middle domain wall should also be smooth. 
This suggests that the ${\bf S}^1$ is fibered over the interval in such a way that 
it degenerates at $R=0$ and $R=L$, resulting in a smooth moduli space with 
topology ${\bf CP}^1$. The K\"ahler metric on this space is squashed compared 
to the round Fubini-Study metric by an amount dependent upon $L,\zeta$ and $m_i$, 
resulting in what is sometimes called the ``sausage model'' \cite{foz}. In a 
similar fashion, one can construct squashed ${\bf CP}^n$ models by 
considering the dynamics of $n$ light domain walls sandwiched between two heavy ones. 
The mirror Landau-Ginzburg theories for these sigma-models were discussed 
in \cite{hv,hk} and are of the $A_n$ affine Toda form. For example, in the 
${\bf CP}^1$ case, the mirror theory has a potential of the form 
$V\sim \exp(-R) + \exp(R-L)$. It is clear that in our effective theory 
approach, such a potential will be generated by the repulsive force between the 
light domain wall and the two heavy bookends. Details will be provided elsewhere.
\begin{figure}[htb]
\begin{center}
\epsfxsize=5.5in\leavevmode\epsfbox{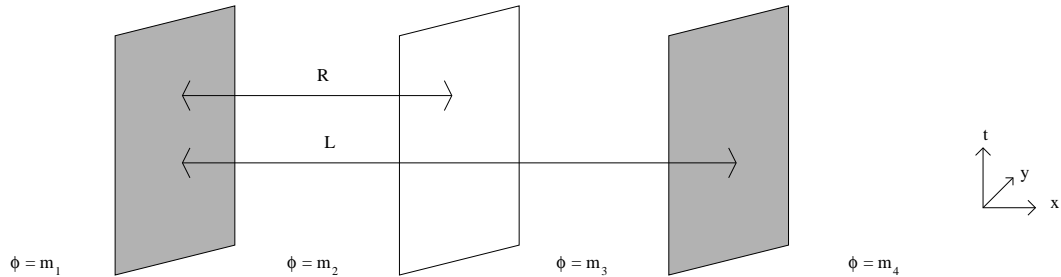}
\end{center}
\caption{Three domain walls. When the outer two become heavy (shown by the 
shading) the light, middle domain wall is restricted to bounce between them. This 
gives a realisation of the ${\bf CP}^1$ sigma-model.}
\label{figure3}
\end{figure}

It is worth noting that, in each of the examples above, the classical 
theory admits domain wall solutions with arbitrary separation while, in 
the low-energy effective theory, the domain walls repel. One 
may wonder if this phenomenon is generic: do quantum effects always 
generate a repulsive force between domain walls in these theories? 
In fact they do not. 
To see this, we may invoke mirror symmetry of three dimensional gauge 
theories. Recall that the $U(1)$ gauge theory with $N$ hypermultiplets 
is equivalent to the $A_{N-1}$ quiver theory, where the Higgs and Coulomb 
branches of the two theories are exchanged \cite{is}. In the quiver theory, 
one may check that domain walls repel classically. However, after 
integrating out the matter multiplets, one finds that the force between 
the walls vanishes. Thus, in this theory, the quantum effects lead to an 
attractive force between the walls which precisely cancels the classical 
repulsive force.
\para
Throughout this paper, we have focused on deriving the quantum duality 
of two-dimensional field theories from classical domain wall 
dynamics. 
However, one could turn the issue around and ask 
if either of the ${\cal N}=(2,2)$ SCFTs described by \eqn{2dbh} or \eqn{lv} 
can teach us about the quantum dynamics of domain walls in the three-dimensional 
theory. For the bosonic CFT, both the spectrum and partition function 
are known \cite{dvv,janandnickandami} and these calculations were recently 
extended to the superconformal case \cite{hh}. As well as the expected continuum of 
scattering states, the spectrum includes towers of discrete states with wavefunctions 
localised at the tip of the black hole. (Unitarity of these representations 
was discussed in \cite{pakman}). In the bosonic theory, these states are 
labeled by their representation under $SL(2,{\bf R})$ and 
their momentum along the $U(1)$ isometry, and include contributions from 
winding modes. From the three-dimensional perspective, the discrete states 
correspond to bound states of domain walls. In particular, the momentum modes of 
the sigma-model are related to dyonic domain walls, known as Q-kinks, which carry a 
global flavour charge \cite{abtown,dw1}. It would  
be of interest to make the mapping between the chiral primaries of the SCFT 
and BPS domain wall states more precise.


\section*{Acknowledgements}

My thanks to Ami Hanany, Kentaro Hori, 
Joe Minahan, Martin Schnabl and especially Jan Troost for useful comments 
and discussions on various aspects of this work. I'd also like to thank 
Jerome Gauntlett, Nick Manton and Paul Townsend for conversations in the 
distant past. 
I'm indebted to Andreas Karch and the University of Washington for their 
kind hospitality and, as always, to the Pappalardo family for their benevolence. 
This work was supported by a Pappalardo fellowship and also in part by funds 
provided by the U.S. Department of Energy (D.O.E.) under cooperative 
research agreement \#DF-FC02-94ER40818.

\end{document}